\documentstyle[12pt,a4wide]{article}

\def\L{{\cal L}}
\def\M{{\cal M}}

\def\C{{\cal C}}
\def\D{{\cal D}}

\def\endproof{\hbox to \hsize{\hfil $\Box$}}

\newtheorem{theorem}{Theorem}
\newtheorem{prop}[theorem]{Proposition}
\newtheorem{lemma}[theorem]{Lemma}
\newtheorem{example}{Example}
\newtheorem{conjecture}{Conjecture}

\begin{document}

\bigskip

\bigskip
\centerline{{\Large\bf Degenerate Frobenius manifolds and the}}

\vspace{.2in}
\centerline{{\Large\bf bi-Hamiltonian structure of rational}}

\vspace{.2in}
\centerline{{\Large\bf Lax equations}}

\vspace{.3in}
\centerline{{\bf I.A.B. Strachan}}\vspace{.1in}
\centerline{Department of Mathematics, University of Hull,}
\vspace{.1in}
\centerline{Hull, HU6 7RX, England.}
\vspace{.1in}
\centerline{e-mail: i.a.b.strachan@hull.ac.uk}

\vspace{.4in}
\centerline{{\bf Abstract}}

\vspace{.3in}
\small
\parbox{5.8in}{The bi-Hamiltonian structure of certain multi-component integrable systems,
generalizations of the dispersionless Toda hierarchy, is studied for systems derived from a
rational Lax function. One consequence of having a rational rather than a polynomial Lax
functions is that the corresponding bi-Hamiltonian structures are degenerate, i.e. the
metric which defines the Hamiltonian structure has vanishing determinant.
Frobenius manifolds provide a natural setting in which to study the bi-Hamiltonian structure
of certain classes of hydrodynamic systems. Some ideas on how this structure may
be extended to include degenerate bi-Hamiltonian structures, such as those given in the first
part of the paper, is given.}
\normalsize

\bigskip

\section{Introduction }

\bigskip

Poisson brackets of hydrodynamic type were introduced by Dubrovin and Novikov in \cite{DN}
where they gave a complete description of Poisson brackets of the form

\begin{equation}
\{u^i(x),u^j(y)\} = g^{ij} [u(x)] \delta^\prime(x-y) + \Gamma^{ij}_k [u(x)] u^k(x) \delta(x-y)
\label{eq:pb}
\end{equation}
under the non-degenerate condition $\det(g^{ij})\neq 0\,.$ This defines a skew-symmetric Poisson
bracket on functionals

\[
\{ I,J\} = \int dx \frac{\delta I}{\delta u^i(x)} \widehat{A^{ij}} 
\frac{\delta J}{\delta u^j(x)}
\]
where
\[
\widehat{A^{ij}} =g^{ij} [u(x)] \frac{d~}{dx} + \Gamma^{ij}_k [u(x)] u^k(x) \,.
\]
The conditions on $g^{ij}$ and $\Gamma^{ij}_k$ necessary in order for
(\ref{eq:pb}) to define a Hamiltonian structure, under the non-degenerate
condition $\det(g^{ij})\neq 0\,,$ have a natural geometric interpretation \cite{DN}:

\begin{theorem} Under the non-degenerate
condition $\det(g^{ij})\neq 0\,$ the bracket (\ref{eq:pb}) defines a
Hamiltonian structure if and only if:

\bigskip

a) ${\bf g}=(g^{ij})^{-1}$ defines a (pseudo)-Riemannian metric;

\medskip

b) $\Gamma^{ij}_{\phantom{ij}k} = - g^{is} \Gamma^{j}_{\phantom{j}sk}\,,$ where
$\Gamma^{j}_{\phantom{j}sk}$ are the Christoffel symbols of the Riemannian connection
defined by ${\bf g}\,;$

\medskip

c) the Riemann curvature tensor of $\bf g$ vanishes.

\end{theorem}

\noindent This result, and its interpretation in terms of differential
geometry, rests on the non-degeneracy condition on the metric. However,
this is not a necessary condition for (\ref{eq:pb}) to define a
Hamiltonian structure and the full result, with no a priori restriction
on $g^{ij}$ was derived by Grinberg \cite{G} and Dorfmann \cite{Dorf}:

\begin{theorem} The bracket (\ref{eq:pb}) defines a
Hamiltonian structure if and only if the pair $(g,\Gamma)$ satisfy the
conditions:  

\bigskip

\begin{eqnarray}
g^{ij} & = & g^{ji} \,; \label{eq:Ga}\\
\frac{\partial g^{ij}}{\partial u^k} & =  & \Gamma^{ij}_k + \Gamma^{ji}_k \,; \label{eq:Gb}\\
g^{ij} \Gamma^{rs}_i & =  & g^{ri} \Gamma^{js}_i \,;\label{eq:Gc}\\
\Gamma^{ij}_t \Gamma^{tk}_r - \Gamma^{ik}_t \Gamma^{tj}_r & = & g^{ti}
\Bigg( \frac{\partial \Gamma^{jk}_r}{\partial u^t} -
\frac{ \partial \Gamma^{jk}_t}{\partial u^r}\Bigg)\label{eq:Gd}
\end{eqnarray}

\noindent and

\begin{equation}
\sum_{ {\rm cyclic~sum~on~} i,j,k} \Bigg[ \Bigg(
\frac{\partial\Gamma^{ij}_t}{\partial u^q} - \frac{\Gamma^{ij}_q}{\partial u^t} \Bigg)
\Gamma^{tk}_r+\Bigg(
\frac{\partial\Gamma^{ij}_t}{\partial u^r} - \frac{\Gamma^{ij}_r}{\partial u^t} \Bigg)
\Gamma^{tk}_q \Bigg] = 0 \,.
\label{eq:Ge}
\end{equation}

\noindent If $\det{g^{ij}}\neq 0$ then the last equation is a consequence of
the earlier equations.

\end{theorem}

\noindent[{\bf N.B.} there is a minor error in \cite{G} in the order of the
indices in equation (\ref{eq:Gc})]. In this more general situation it is not possible
to give a clear geometric interpretation of these equations. They
define an integrable distribution but their differential geometric content is less
clear.
One can define a covariant derivative like object

\[
\nabla^i \xi^j = \partial^i \xi^j - \Gamma^{ij}_k \xi^k\,,
\]
where $\partial^i = g^{ij} \partial_j\,,$ with the property (when suitably
extended to tensors) that $\nabla^i g^{jk}=0\,,$ though the \lq connection\rq~cannot
defined in terms of the \lq metric\rq. With such
a covariant derivative one can introduce a \lq curvature\rq~by the equation

\[
(\nabla^r \nabla^s - \nabla^s \nabla^r )\xi^t = - R^{rst}_k \xi^k
\]
and the third equation above is now just the vanishing of this curvature.
Such a description is not very natural; one cannot lower indices and the
interpretation of the last equation remains unclear. However the terms
\lq metric\rq~and \lq connection\rq~will be used to denote these objects,
and a pair satisfying these equations will be called a $(g,\Gamma)$-pair.

\bigskip

The purpose of this paper is to study the bi-Hamiltonian structure of dispersionless
integrable systems defined by the Lax equation
\footnote{The variables $\tau_n$ will be used to denote the times,
$t$ being reserved for flat coordinates in which the components $\eta^{ij}$ are constants.}

\begin{equation}
\frac{\partial \L}{\partial \tau_n} = \{ ( \L^{\frac{n}{N-M}} )_{+}, \L\}_{PB}
\label{eq:lax}
\end{equation}

\noindent where $\{ f,g\}_{PB} = p (\partial_p f \partial_x g - \partial_x f \partial_p g)\,,$
$\L$ is given by a rational function

\[
\L=\frac{\rm polynomial~of~degree~N}{\rm polynomial~of~degree~M}
\]
with the single constraint $N>M\,,$
and $(~)_{+}$ denotes the projection onto non-negative powers of $p$ under a formal expansion
in powers of $p\,.$
In an earlier paper \cite{FS} this system was studied but a complete description of the
Hamiltonian structure
was not given. The simplest example of such a system is the continuum Toda equations

\begin{equation}
\begin{array}{rcl}
S_\tau & = & P_x \,, \\
P_\tau & = & P S_x\,.
\end{array}
\label{eq:example}
\end{equation}

\noindent which is generated for the above Lax equation (\ref{eq:lax}) with a Lax function

\[
\L = p^2 + S(x,t) + \frac{P(x,t)}{p}\,.
\]
This paper aims to extend these earlier results from polynomial Lax functions to
rational Lax functions and to relate these results to the theory of Frobenius manifolds
\cite{D}. It will turn out that in the rational case the Hamiltonian structure
is degenerate, so the more general description of Grinberg and Dorfmann will have to be
utilised to give a complete description of the bi-Hamiltonian structure of the hierarchy.
This in turn implies that a new concept of a degenerate Frobenius manifold is required.

\bigskip

In the next section a summary of the pertinent result of \cite{FS} will be given, and this will
also serve to fix the notation used. Full details will not be given and the reader should
consult the earlier paper for the proofs. In section 3 the polynomial case will be
studied in more detail (and this will relate the results of \cite{FS} to more recent work of
Dubrovin and Zhang \cite{DZ}) before the full rational case is studied in section 4.
The properties of a degenerate Frobenius manifold is introduced by way of an extended example
in section 5.

\bigskip

Throughout this paper various different coordinate systems will be used, and the
resulting transformations from one system to another will be important. The notation
$g^{ij}(s)$ will be used to denote the components of the metric in the $s^i$-coordinate
system, so the transformation from $s^i$ to $t^i$ coordinates will be
written

\[
g^{ij}(t) = \frac{\partial t^i}{\partial s^p}\frac{\partial t^j}{\partial s^q} g^{pq}(s)\,,
\]

\noindent rather than using different founts and alphabets for the different coordinate
systems.

\bigskip

\section{Conservation Laws and Evolution Equations}

In order to study rational functions it is convenient, and indeed necessary in order
to obtain some results, to factorize the numerator and denominator of the rational
function, so

\begin{eqnarray*}
\L & = & \frac{ \prod_{i=1}^N (p+u^i) }{ \prod_{i=N+1}^{N+M} (p+u^i) } \,, \\
& = & \prod_{i=1}^{N+M} (p+u^i)^{\varepsilon_i} \,.
\end{eqnarray*}
Here it will be assumed that $\varepsilon_i=\pm 1$ and that the numerator and denominator
have no common root. With these conditions and $N>M$ the Lax function is of the general form

\[
\L ={\rm polynomial~of~degree~}(N-M) ~ +
\sum_{i=N+1}^{N+M} {\rm simple~poles}\,.
\]
Such a factorization of the Lax function was introduced by Kupershmidt \cite{K} (though this could
also be viewed as a Viet\'e transformation) and the variables $u^i$ will be called
modified variables. One advantage of such a factorization is that it puts all the
fields on an egalitarian footing, i.e. the permutation group $S^N$ acts on the zeros of $\L$
and the permutation group $S^M$ acts on the roots of $\L\,,$ and this
drastically reduces the complexity of the calculations.

\bigskip

The flows are given by the Lax equation (\ref{eq:lax}) which may be calculated
explicitly

\begin{equation}
u^i_{\tau_n} = A_i^{(n)} u^i_x + \sum_{j \neq i} u^i B_{ij}^{(n)} u^j_x
\label{eq:rationalflow}
\end{equation}

\noindent where

\[
A_i^{(n)}=\Big( \frac{\epsilon_i n}{N-M}-1\Big)
\sum_{ \{ r_j\,: \sum_{j=1}^{N+M} r_j= n \} }
\Bigg[
\prod_{ \scriptstyle k=1 \atop \scriptstyle k\neq i}^{N+M}
\pmatrix{ \frac{\epsilon_k n}{N-M} \cr r_k } (u^k)^{r_k}
\Bigg]
\pmatrix{ \frac{\epsilon_i n}{N-M} -2 \cr r_i - 1} (u^i)^{r_i}
\]

\noindent and $B_{ij}^{(n)} = $

\[
\frac{\epsilon_j n}{N-M}
\sum_{ \{ r_j\,: \sum_{j=1}^{N+M} r_j= n-1 \} }
\Bigg[
\prod_{ \scriptstyle k=1 \atop \scriptstyle k\neq i,j}^{N+M}
\pmatrix{ \frac{\epsilon_k n}{N-M} \cr r_k } (u^k)^{r_k}
\Bigg]
\pmatrix{ \frac{\epsilon_i n}{N-M} -1 \cr r_i } (u^i)^{r_i}
\pmatrix{ \frac{\epsilon_j n}{N-M} -1 \cr r_j } (u^j)^{r_j}\,.
\]
Care has to be taken in evaluating the binomial coefficients for negative and fractional
numbers. These must be interpreted in terms of $\Gamma$-function, so

\[
\pmatrix{a \cr b} = \frac{  \Gamma(a+1) }{ \Gamma(a-b+1) \Gamma(b+1) }\,.
\]

\noindent It also follows from the proof of these results (though not explicitly
mentioned in \cite{FS}) that

\begin{eqnarray*}
\C & = & \left. \L \right|_{p=0} \,, \\
& = & \prod_{i=1}^{M+N} (u^i)^{\epsilon_i}
\end{eqnarray*}
is a independent of all the times, i.e.
\[
\frac{\partial \C}{\partial \tau_n} = 0 \quad\quad n=1\,, \ldots \,, \infty\,.
\]
The functions $\C$ will turn out to be a Casimir for the bi-Hamiltonian structure of this
hierarchy.

\bigskip

\noindent Conservation laws are similarly defined, the conserved charges being given by

\begin{equation}
Q^{(n)} = \frac{1}{2\pi i} \oint \L^{ \frac{n}{N-M} } \frac{dp}{p}\,.
\label{eq:hamdensity}
\end{equation}

\noindent These may be derived explicitly

\[
Q^{(n)} = \sum_{ \{ r_i \,: \sum_{i=1}^{N+M} r_i= n \} }
\Bigg\{  \prod_{i=1}^{N+M}
\pmatrix{ \frac{\epsilon_i n}{N-M} \cr r_i } (u^i)^{r_i} \Bigg\}\,.
\]

\noindent Under a suitable change of variable, these polynomials take the
form of generalised hypergeometric functions, a result which remains to be
exploited. The corresponding functionals

\begin{equation}
H^{(n)} = \int Q^{(n)}\, dx
\label{eq:hamiltonians}
\end{equation}
will turn out to be the Hamiltonians of the system (\ref{eq:rationalflow}).

\bigskip

\section{Polynomial Lax equations}

In this section the bi-Hamiltonian structure of the hierarchy defined by a
Lax function

\[
\L = p^{-M} \prod_{i=1}^{N} (p+u_i)\,, \qquad 0<M<N\,,
\]
will be derived, this generalising the results of \cite{FS} where the special case
$M=1$ was studied. Having derived one Hamiltonian structure, the intersection form
in the language of Frobenius manifold, one may use a result of Dubrovin to find a second
compatible Hamiltonian structure.

\begin{prop}

The Hamiltonian structure of the hierarchy defined by equation (\ref{eq:lax}) is
given by the non-degenerate metric

\begin{equation}
g^{ij}(u) = \cases{ [1-(N-M)] u^i u^i & if $i=j\,,$ \cr u^i u^j & if $i\neq j\,.$}
\label{eq:polymetric}
\end{equation}

\end{prop}

\noindent{\bf Comment} This is clearly a flat, non-degenerate metric, and so defines
a Hamiltonian structure. What is less clear is whether this structure, coupled to the
Hamiltonians given by (\ref{eq:hamiltonians}), gives rise to the flows
defined by (\ref{eq:lax}). This
may be shown to be the case by direct calculation. An alternative proof, viewing the
polynomial as a reduction of the rational case, will follow from the Theorem 7 in
section 4.

\bigskip

A bi-Hamiltonian structure is more than just two Hamiltonian structures; the
two structures $\{~,~\}_1$ and $\{~,~\}_2$ have to be compatible,
i.e. $\{~,~\}=\{~,~\}_1+\lambda \{~,~\}_2$
must be a Hamiltonian structure for all values of $\lambda\,. $
For non-degenerate Poisson brackets of
hydrodynamic type this compatibility condition implies that, for arbitrary $\lambda\,:$

\bigskip

(a) the metric $g^{ij} = g_1^{ij} + \lambda g_2^{ij}$ is flat (such a metric is sometimes
referred to as a flat pencil); 

\bigskip

(b) the metric connection for this metric has the form
$\Gamma^{ij}_k = \Gamma^{ij}_{1k} + \lambda\Gamma^{ij}_{2k}\,.$ 

\bigskip

\noindent A result of Dubrovin \cite{D} (actually a special case of a more general result of
Magri \cite{Magri}) will enable the bi-Hamiltonian structure to be
found:

\begin{lemma}

If for a flat metric in some coordinate system $x^1\,,\ldots\,, x^n$ both the
components $g^{ij}(x)$ of the metric and $\Gamma^{ij}_k(x)$ of the corresponding
metric connection depend linearly on the coordinate $x^\bullet$ then the metrics

\begin{eqnarray*}
g^{ij}_1 & = & g^{ij}\,,\\
g^{ij}_2 & = & \partial_\bullet g^{ij}
\end{eqnarray*}
form a flat pencil, under the assumption that $\det[g^{ij}_2]\neq 0\,.$ The
corresponding metric connection has the form
\begin{eqnarray*}
\Gamma^{ij}_{1k} & = & \Gamma^{ij}_k\,, \\
\Gamma^{ij}_{2k} & = &\partial_\bullet \Gamma^{ij}_k\,.
\end{eqnarray*}

\end{lemma}

\noindent The proof of this result is straightforward, and an alternative proof to that
given in \cite{D} will follow from a result given in the next section
where this lemma is extended
to degenerate Hamiltonian structures.

\bigskip In order to find such a coordinate system it is necessary to perform a number
of coordinate transformations on the metric (\ref{eq:polymetric}). This will be achieved
in two stages. First define variables \cite{DZ}

\begin{eqnarray*}
z^1 & = & + x^1 \,, \\
z^i & = & +x^i - x^{i-1} \,, \qquad\qquad i=1\,,\ldots \,, N-1\,,\\
z^N & = & -x^{N-1}
\end{eqnarray*}
(so $\sum_{i=1}^N z^i = 0$) and then

\[
u^i = e^{ \frac{1}{N}x^N - z^i}\,, \qquad\qquad i=1\,,\ldots \,, N\,.
\]
Such a coordinate transformation has a nature interpretation in terms of the
Weyl group $W(A_{N-1})\,,$ which act by permutation of the coordinates $z^i$ on
the hyperplane $\sum_{i=1}^N z^i = 0$. In this $x$-coordinate systems
the components of the metric become, up to an overall factor of $(M-N)\,:$

\begin{eqnarray*}
g^{ij}(x) & = &
\left(
\begin{tabular}{ccccc|c}
 2 & -1 &  0 & \ldots & 0 & 0\\
-1 &  2 & -1 & \ldots & 0 & 0 \\
 0 & -1 &  2 & \ldots & 0 & 0 \\
\vdots & \vdots & \vdots &  & \vdots & \vdots \\
 0 &  0 &  0 & \ldots & 2 & 0\\
\cline{1-6}
 0 &  0 &  0 & \ldots & 0 & $-\frac{M}{N(N-M)}$
\end{tabular}
\right)\\
& & \\
& = & \left(
\begin{tabular}{c|c}
\begin{tabular}{c} Cartan matrix \\ of $A_{N-1}$ \end{tabular} & 0 \\
\cline{1-2}
 0 & $-d_M^{-1}$
\end{tabular}
\right)\,.
\end{eqnarray*}
The final entry is defined naturally using the Weyl group structure on
$A_{N-1}\,,$

\begin{eqnarray*}
d_M & = & \frac{M(N-M)}{N} \,,\\
& = & (\omega_M\,,\omega_M)
\end{eqnarray*}
where $(~,~)$ is the Euclidean inner product and $\omega_i$ are the fundamental weights
\cite{DZ}.

\bigskip

What these coordinate transformation show is that the Hamiltonian structure
coincides with those found by Dubrovin and Zhang, so their results may be
used to complete the second part of this arguments. In particular they
show that in terms of the symmetric functions

\begin{eqnarray*}
s^1 & = & \sum_i u^i \,, \\
s^2 & = & \sum_{i<j} u^i u^j \,, \\
\vdots &  & \vdots \\
& & \\
s^N & = & \prod_i u^i \,.
\end{eqnarray*}

\noindent the metric (\ref{eq:polymetric}) will be linear in the variable
$s^M\,,$ and hence lemma 4 may be used to find the bi-Hamiltonian structure.
The Jacobian of this transformation from modified to the original variables
is just the Vandemonde determinant,

\begin{eqnarray*}
\frac{\partial(s^1\,,\ldots \,, s^N)}{\partial(u^1\,, \ldots \,, u^N)}
& = &
\left |
\begin{array}{cccc}
1 & 1 & \ldots & 1 \\
\sum_{i\neq 1} u^i & \sum_{i\neq 2} u^i & \ldots & \sum_{i \neq N} u^i \\
\vdots & \vdots & \ddots & \vdots \\
\prod_{i\neq 1} u^i & \prod_{i\neq 2} u^i & \ldots & \prod_{i \neq N} u^i
\end{array}
\right | \\
& & \\
& = & \prod_{i<j} (u^i - u^j)\,.
\end{eqnarray*}
This defines the discriminant hypersurface, a caustic, on which $\L$ has
multiple roots. By assumption $\varepsilon_i = \pm 1$ so all the roots are simple
and hence the fields are well defined away from this surface. Hence \cite{DZ}:

\begin{lemma}

The metric (\ref{eq:polymetric}), when written in terms of the symmetric
variables $s^i$ is linear in the variable $s^M\,.$

\end{lemma}

Before performing these calculations one should note that in terms of these
symmetric variables
the Lax function takes the more familiar form

\[
\L = p^{-M} \Big[ p^N + p^{N-1} s^1 + \ldots + s^N \Big]\,,
\]
these symmetric variables coinciding with the original, unmodified
variables. It also follows from the Lax equation (\ref{eq:lax}) that
the variable $s^M$ is special for another reason, namely it is the single variable
for which the conserved charges $Q^{(n)}$ obey the relation

\[
Q^{(n-1)} = {\rm constant~} \frac{\partial Q^{(n)}}{\partial s^M}\,.
\]

\bigskip

\begin{prop}

The first Hamiltonian structure, in terms of the modified variables, is given by

\begin{eqnarray*}
\eta^{ij}(u) & = & \L_{\frac{\partial~}{\partial s^\bullet}} g^{ij}(u) \,, \\
& = &\frac{\partial{\phantom{S}}}{\partial s^\bullet} g^{ij} -
\frac{\partial \alpha^i_\bullet}{\partial u^k} g^{kj} -
\frac{\partial \alpha^j_\bullet}{\partial u^k} g^{ik}
\end{eqnarray*}
where the functions $\alpha^i_\bullet(u)$ are defined by

\[
\frac{\partial{\phantom{s^\bullet}}}{\partial s^\bullet}=
\alpha^i_\bullet(u) \frac{\partial{\phantom{u^i}}}{\partial u^i}\,
\]
and $\L_{\frac{\partial~}{\partial s^\bullet}}$ is the Lie derivative along the
vector field $\frac{\partial~}{\partial s^\bullet}\,.$

\end{prop}

\bigskip

\noindent{\bf Proof} The transformation between the modified variables and the
symmetric variables induces the transformation

\[
\left(
\begin{array}{c}
\frac{\partial~}{\partial u^1} \\
\vdots \\
\frac{\partial~}{\partial u^N}
\end{array}
\right)
=
\left(
\begin{array}{cccc}
1 & \sum_{j\neq 1} u^j & \ldots & \prod_{j\neq 1} u^j \\
\vdots & \vdots & \ddots & \vdots \\
1 & \sum_{j\neq N} u^j & \ldots & \prod_{j\neq N} u^j
\end{array}
\right)
\left(
\begin{array}{c}
\frac{\partial~}{\partial s^1} \\
\vdots \\
\frac{\partial~}{\partial s^N}
\end{array}
\right)
\]
and hence by inverting the Vandemonde determinant

\[
\frac{\partial{\phantom{s^\bullet}}}{\partial s^\bullet}=
\alpha^i_\bullet(u) \frac{\partial{\phantom{u^i}}}{\partial u^i}\,,
\]
this defining the functions $\alpha^i_\bullet(u)\,.$ By lemmas 4 and 5 it follows,
by starting with the metric (\ref{eq:polymetric}) in the $u^i$-variables, transforming
to the $s^i$-variables, differentiating with respect to $s^\bullet$ and then transforming
back to the $u^i$-variables that

\[
\eta^{ij}(u)=\frac{\partial u^i}{\partial s^m}\frac{\partial u^j}{\partial s^n}
\frac{\partial~}{\partial s^\bullet} \left[
\frac{\partial s^m}{\partial u^r}
\frac{\partial s^n}{\partial u^s}g^{rs}(u) \right]
\]
is the required flat metric, which defines the second Hamiltonian structure. Expanding
yields
\[
\eta^{ij}(u) = \frac{\partial~}{\partial s^\bullet} g^{ij}  + 
\frac{\partial u^i}{\partial s^m} \left(
\frac{\partial~}{\partial s^\bullet} \frac{\partial s^m}{\partial u^r}\right) g^{rj}
 + 
\frac{\partial u^j}{\partial s^n} \left(
\frac{\partial~}{\partial s^\bullet} \frac{\partial s^n}{\partial u^s}\right) g^{is}\,.
\]
But
\begin{eqnarray*}
\left[\frac{\partial~}{\partial s^\bullet}\,,\frac{\partial~}{\partial u^k}\right]& = &
\left[ \alpha^i_\bullet \frac{\partial~}{\partial u^i}\,,\frac{\partial~}{\partial u^k}
\right]\,,\\
& = & - \frac{\partial \alpha^j_\bullet}{\partial u^k} \, \frac{\partial~}{\partial u^j}\,.
\end{eqnarray*}
This, together with the definition of the Lie derivative and 
\[
\frac{\partial s^\beta}{\partial s^\bullet} = \delta^\beta_\bullet
\]
yields the result.

\bigskip

\endproof

\noindent This proposition is just an application of Magri's more
general result \cite{Magri}\,.

\bigskip

\begin{example}

\bigskip

For arbitrary $N$ and $M=1$ the distinguished coordinate is $s^N$ (so $\bullet=N$ in the
above formulae) and the
functions $\alpha^i_N$ are

\[
\alpha^i_N = \frac{u^i}{\prod_{r\neq i}(u^N-u^r)}
\]
and hence one may calculate $\eta^{rs}$ explicitly:

\begin{eqnarray*}
\eta^{ij}(u) & = & (N-1)\frac{ u^i u^j}{u^i-u^i} \,[\alpha^i_N - \alpha^j_N]\,,
\quad\quad r\neq s\,,\\
\eta^{ii}(u) & = & 2 (N-1) \frac{ (u^i)^2}{\prod_{k \neq i} (u^i - u^k)}
\left[ 1 - u^i \sum_{n\neq i} \frac{1}{u^i - u^n} \right]\,.
\end{eqnarray*}
This, togther with (\ref{eq:polymetric}), constitutes the bi-Hamiltonian structure
for the hierarchy (\ref{eq:lax}), also known as the continuum Toda hierarchy.

\bigskip

If $N=2$ then

\[
\eta^{ij}(u) = \frac{uv}{(u-v)^2}\left(
\begin{array}{cc}
-2u & u+v \\
u+v & -2v
\end{array}\right)
\]

\end{example}

\noindent This example also shows an interesting result of the
transformation from the original
to the modified variables; in the original variables the form of $g^{ij}$ is more
complicated than the form of $\eta^{ij}$ while in the modified variables the complexities
are interchanged.

\bigskip

These results depend crucially on the properties of $d_M\,.$ To see this consider
the flat metric 

\[
h^{ij}(u,v,w)=
\left(
\begin{array}{ccc}
a u^2 & u v & u w\\
u v & a v^2 & v w\\
u w & v w & a w^2
\end{array}
\right)\,,
\]
this being (\ref{eq:polymetric}) with $N=3$ and $1-(N-M)$ replaced with an
arbitrary constant $a\,.$ We assume that this metric is invertible (so
$a \neq 1\,,-2\,)\,.$
In terms of symmetric variables $S=u+v+w\,,P=u v + v w + w u$ and $Q=u v w$ this
takes the form

\[
h^{ij}(S,P,Q)=
\left(
\begin{array}{ccc}
a s^2 + 2(1-a) P & (1+a) S P + 3(1-a) Q & (2+a) S Q \\
(1+a) S P + 3(1-a) Q & 2(1+a) P^2 + 2(1-a) S Q & 2(2+a) P Q \\
(2+a) S Q & 2(2+a) P Q & 3 (2+a) Q^2
\end{array}
\right)\,.
\]

\noindent For general values of $a$ the entries are not linear in any of the
variables. The metric cannot depend on $Q$ linearly as this would imply
$a=-2\,.$ For the entries to depend linearly on $Q$ would imply $a=-1$ and this
corresponds to (\ref{eq:polymetric}) with $M=1\,.$ 
For the entries to depend linearly on $S$ would imply $a=0$ and this
corresponds to (\ref{eq:polymetric}) with $M=2\,.$ Thus any requirement that
the metric depends linearly on one of the symmetric variables forces the
metric to take one of the above known forms. Of course this does not rule out the
possibility that in some other coordinate systems the components of the
metric do become linear in some variable.

\bigskip

\section{Rational Lax equations}

In this section the evolution equations (\ref{eq:rationalflow}) will be written in
Hamiltonian form. The resulting Hamiltonian structure turns out to be degenerate, so
the results of Dubrovin used in the last section to derive the bi-Hamiltonian structure
cannot be used without modification. These modifications turn out to be minor and a
version of lemma 4 will hold for degenerate Hamiltonian systems.

\begin{theorem}
(A) In terms of the variables ${\tilde u}^i=\log u^i$ the evolution
equations (\ref{eq:rationalflow}) may be
written in Hamiltonian form

\[
{\tilde u}^i_{\tau_n} = \sum_j m^{ij} \D \left( \frac{ \delta H^{(n)}}{\delta {\tilde u}^j}\right)
\]
where $m^{ij}$ is the constant matrix

\begin{equation}
m^{ij}=\left(
\begin{array}{ccccc}
\alpha_1 & 1 & 1 & \ldots & 1 \\
1 & \alpha_2 & 1 & \ldots & 1 \\
1 & 1 & \alpha_3 & \ldots & 1 \\
\vdots & \vdots & \vdots & \ddots & \vdots \\
1 & 1 & 1 & \ldots & \alpha_{N-M}
\end{array}\right)\,,
\label{eq:defm}
\end{equation}

\noindent $\alpha_i = 1 - \varepsilon_i(N-M)$ and 

\[
H^{(n)} = \int dx \,\sum_{ \{ r_i \,: \sum_{i=1}^{N+M} r_i= n \} }
\Bigg\{  \prod_{i=1}^{N+M}
\pmatrix{ \frac{\epsilon_i n}{N-M} \cr r_i } e^{r_i {\tilde u}^i} \Bigg\}\,
\]

\noindent (B) In terms of the original variables the $(g,\Gamma)$ pair

\begin{eqnarray*}
g^{ij}(u) & = & m^{ij} u^i u^j \,, \\
\Gamma^{ij}_k(u) & = & \delta^j_k m^{ij} u^i
\end{eqnarray*}
define a degenerate Hamiltonian structure, satisfying the conditions of
Theorem 2.

\end{theorem}

\bigskip

{\noindent{\bf Proof}}

\medskip

(A) In terms of the $u^i$ variables the system

\begin{equation}
{\tilde u}^i_{\tau_n} = \sum_j m^{ij} \D \left( \frac{ \delta H^{(n)}}{\delta {\tilde u}^j}\right)
\label{eq:ham}
\end{equation}

\noindent becomes

\[
u^i_{\tau_n} = \sum_i\alpha_i u^i \D\left( u^i \frac{\delta H^{(n)}}{\delta u^i}\right)+
\sum_{j\neq i}  u^i \D\left( u^j \frac{\delta H^{(n)}}{\delta u^j}\right)\,.
\]

\noindent Expanding this yields

\begin{eqnarray*}
u^i_{\tau_n} & = & \phantom{\sum_{j\neq i}}
\left[ \frac{1}{n} \sum_{ \{ r_i \,: \sum_{i=1}^{N+M} r_i= n \} }
[\alpha_i r_i^2 + r_i(n-r_i)]
\prod_{k=1}^{N+M}
\pmatrix{ \frac{\epsilon_k n}{N-M} \cr r_k }(u^k)^{r_k}\right] u^i_x + \\
& &\sum_{j\neq i} \left[ \frac{1}{n} \sum_{ \{ r_i \,: \sum_{i=1}^{N+M} r_i= n \} }
[\alpha_i r_i r_j + r_j(n-r_i)] (u^j)^{-1}\prod_{k=1}^{N+M}
\pmatrix{ \frac{\epsilon_k n}{N-M} \cr r_k }(u^k)^{r_k}\right] u^j_x\,.
\end{eqnarray*}
Using $ \alpha_i = 1 - \varepsilon_i(N-M)$ and various binomial identities
reduces this to

\[
u^i_{\tau_n} = A_i^{(n)} u^i_x + \sum_{j \neq i} u^i B_{ij}^{(n)} u^j_x
\]

\noindent where

\[
A_i^{(n)}=\Big( \frac{\epsilon_i n}{N-M}-1\Big)
\sum_{ \{ r_j\,: \sum_{j=1}^{N+M} r_j= n \} }
\Bigg[
\prod_{ \scriptstyle k=1 \atop \scriptstyle k\neq i}^{N+M}
\pmatrix{ \frac{\epsilon_k n}{N-M} \cr r_k } (u^k)^{r_k}
\Bigg]
\pmatrix{ \frac{\epsilon_i n}{N-M} -2 \cr r_i - 1} (u^i)^{r_i}
\]

\noindent and $B_{ij}^{(n)}=$

\[
\frac{\epsilon_j n}{N-M}
\sum_{ \{ r_j\,: \sum_{j=1}^{N+M} r_j= n-1 \} }
\Bigg[
\prod_{ \scriptstyle k=1 \atop \scriptstyle k\neq i,j}^{N+M}
\pmatrix{ \frac{\epsilon_k n}{N-M} \cr r_k } (u^k)^{r_k}
\Bigg]
\pmatrix{ \frac{\epsilon_i n}{N-M} -1 \cr r_i } (u^i)^{r_i}
\pmatrix{ \frac{\epsilon_j n}{N-M} -1 \cr r_j } (u^j)^{r_j}\,,
\]

\noindent that is, to the equations obtained from the Lax equation (\ref{eq:lax}).
Hence the result.

\bigskip

(B) Rewritting (\ref{eq:ham}) in terms of a $(g,\Gamma)$ pair yields

\begin{eqnarray*}
g^{ij}(u) & = & m^{ij} u^i u^j \,, \\
\Gamma^{ij}_k (u) & = & \delta^j_k m^{ij} u^i
\end{eqnarray*}

\noindent The above argument does not show that the pair $(g,\Gamma)$ defines a
Hamiltonian structure, as the corresponding bracket (\ref{eq:pb}) must define
a Hamiltonian structure for all functionals, not just the specific functionals
used above. In order to show that this pair does define such a structure one
must verify that the equations (\ref{eq:Ga}-\ref{eq:Ge}) hold.
This is entirely straightforward so the
details will be omitted. The degeneracy of the metric follows from
the result, easily proved using elementary row and column operations, that

\[
\det\left(
\begin{array}{cccc|ccc}
1-a & 1 & \ldots & 1 & 1 & \ldots & 1 \\
1 & 1-a & \ldots & 1 & 1 & \ldots & 1 \\
\vdots & \vdots & & \vdots & \vdots & & \vdots \\
1 & 1 & \ldots & 1-a & 1 & \ldots & 1 \\
\cline{1-7}
1 & 1 & \ldots & 1 & 1+a & \ldots & 1 \\
\vdots & \vdots & & \vdots & \vdots & & \vdots \\
1 & 1 & \ldots & 1 & 1 & \ldots & 1+a
\end{array}
\right) = (-1)^N a^{N+M-1} [ a - (N-M)]
\]
where the diagonal blocks are $N\times N$ and $M \times M$ matrices. For the matrix
$m_{ij}$ $ a = N-M$ (since $\alpha_i = 1 - \varepsilon_i(N-M)$) and hence
$\det (g^{ij})=0\,.$ It also follows from these operations that
${\rm rank}(g^{ij})=(N+M)-1\,.$

\endproof

\bigskip

This also shows that this system is only mildly degenerate; the coordinate
transformation that
reduces $g^{ij}$ to a metric with constant entries simultaneously reduce the $\Gamma^{ij}_k$
to zero. For a degenerate metric this need not be the case, and some non-zero $\Gamma^{ij}_k$
can remain \cite{G}.

\bigskip

\begin{lemma}

Let the pair $(g,\Gamma)$ define a degenerate Hamiltonian structure.
If the components of the pair $(g,\Gamma)$ in some coordinate system
$x^1\,,\ldots\,, x^n$ depend linearly on the coordinate $x^\bullet$ then the pair

\begin{equation}
(g + \lambda \partial_\bullet g,
\Gamma + \lambda \partial_\bullet \Gamma)
\label{eq:degenerate}
\end{equation}defines a degenerate Hamiltonian structure for all values of $\lambda\,.$ Hence one
obtains a degenerate bi-Hamiltonian structure.

\end{lemma}

\bigskip

\noindent{\bf Proof}  All that is required is to show that the pair (\ref{eq:degenerate})
satisfies the conditions of theorem 2, given that the original $(g,\Gamma)$ pair\,. This
is straightforward, the first two conditions being trivial. Consider, for example,
condition (\ref{eq:Gc})
in theorem 2:

\[
[(\Gamma^{ij}_k + \lambda_\bullet \Gamma^{ij}_t)(g^{tk}+\lambda\partial_\bullet g^{tk})-
(k\leftrightarrow i)]
 =  \left( 1 + \lambda\partial_\bullet + \frac{\lambda^2}{2} \partial_\bullet^2\right)
[(\Gamma^{ij}_k  - (k\leftrightarrow i)]\,,
\]
this following from that fact that if $g$ and $\Gamma$ depend linearly on
$x^\bullet$ then
\[
\partial_\bullet^2 (\Gamma g) = 2 \partial_\bullet \Gamma\, \partial_\bullet g\,.
\]
Hence if $(g,\Gamma)$ satisfies condition (\ref{eq:Gc}), so does (\ref{eq:degenerate}).
The remaining
conditions are all quadratic in $g$ and $\Gamma$ and so the proof is identical.

\endproof

\bigskip

One may perform a similar sequence of coordinate transformation to those in section 3. Explicitly
let

\begin{eqnarray*}
z^1 & = & + x^1 \,, \\
z^i & = & +x^i - x^{i-1} \,, \qquad\qquad i=1\,,\ldots \,, N\,,\\
z^N & = & -x^{N-1}\,,\\
z^{N+1} & = & + x^{N+1} \,,\\
z^i & = & +x^i - x^{i-1} \,, \qquad\qquad i=N+1\,,\ldots \,, N+M\,,\\
z^{N+M} & = & - x^{N+M-1}
\end{eqnarray*}
and
\[
u^i = e^{ \frac{1}{N}x^N - z^i}\,, \qquad\qquad i=1\,,\ldots \,, N+M\,.
\]
After a permutation is the labels the metric becomes (up to an overall factor)

\[
g^{ij}(x) = 
\left(
\begin{tabular}{c|c|c}
$+\left(\begin{tabular}{c} Cartan matrix \\ of $A_{N-1}$ \end{tabular}\right)$ & 0 & 0   \\
\cline{1-3}
0 &$-\left(\begin{tabular}{c} Cartan matrix \\ of $A_{N+M-1}$ \end{tabular}\right)$ & 0 \\
\cline{1-3}
0 & 0 & - $d_M \left(\begin{tabular}{cc} 1 & 1 \\ 1 & 1 \end{tabular}\right)$
\end{tabular}
\right)
\]
This is considerable scope for the investigation of bi-Hamiltonian structures based on
such block decompositions.

\bigskip

Having derived one Hamiltonian structure it is necessary, before the above lemma can be
applied, to find a suitable coordinate system in which the metric given in theorem 7
becomes linear
in one of the coordinates. This will be done only for the $M=1$ case, i.e. a rational
Lax function with a single pole. Extending these results to an arbitrary number of pole
presents certain problems, which will be discussed later.

\bigskip

The new variable $s^i$ are defined by the following expansion of the rational function
$\L\,,$

\begin{eqnarray*}
\L & = & {\rm polynomial~of~degree~}(N-1) + \frac{\rm function}{p+{\rm pole}}\,,\\
& = & \sum_{n=0}^{N-1} p^n s^{N-1-n} + \frac{s^N}{p+s^{N+1}}\,.
\end{eqnarray*}

\noindent To express the $s^i$ as functions of the variables $u^i$ it is convenient
to introduce the basic symmetric functions of the variables $u^1\,,\ldots\,,u^N\,:$

\[
\sigma^0 = 1 \,,\quad
\sigma^1 = \sum_i u^i \,, \quad
\sigma^2 = \sum_{i<j} u^i u^j \,, \quad
\ldots \quad
\sigma^N = \prod_i u^i \,,
\]

\noindent so

\[
\prod_{i=1}^N (p+u^i) = \sum_{i=0}^N p^i \sigma^{N-i}\,.
\]

\noindent By expanding the various expressions for $\L$ one obtains

\begin{eqnarray}
s^0 & = & 1 \,,\nonumber\\
s^r & = & \sum_{n=0}^r (-1)^n \sigma^{r-n} (u^{N+1})^n \,,\quad\quad r=1\,,\ldots\,,N\,,
\label{eq:newvar}\\
s^{N+1} & = & u^{N+1}\nonumber\,.
\end{eqnarray}
It is in these variables that the pair $(g,\Gamma)$ will become linear in one of the
variables.

\bigskip

\begin{example} For $N=3\,,M=1$

\begin{eqnarray*}
\L & = &\frac{(p+u)(p+v)}{p+w}\,, \\
& = & p + (u+v-w) + \frac{(u-w)(v-w)}{p+w}\,,
\end{eqnarray*}
and hence
\begin{eqnarray*}
s^1 & = & u+v-w \,,\\
s^2 & = & uv-w(u+v)+w^2\,,\\
s^3 & = & w\,,
\end{eqnarray*}
in accordance with (\ref{eq:newvar}).

\end{example}

\bigskip

\noindent The following result will be required in the next theorem:

\bigskip

\begin{lemma}

With the variables $s^m$ defined above,

\[
\frac{\partial s^m}{\partial u^i}+\frac{ \partial s^m}{\partial u^{N+1}} =
(u^{N+1} - u^i) \{ {\rm polynomial~of~degree~}\alpha-2\}\,,\quad\quad m\,,i=1\,,\ldots\,
N\,.
\]

\end{lemma}

\bigskip

\noindent{\noindent\bf Proof}

One may write $\sigma^n$ as $\sigma^n = u^i {\tilde\sigma}^{n-1}+{\tilde\sigma}^n$ so
\[
\frac{\partial \sigma^n}{\partial u^i} = {\tilde\sigma}^{n-1}\,.
\]
Hence
\begin{eqnarray*}
\left.\frac{\partial s^m}{\partial u^{N+1}}\right|_{u^{N+1}=u^i} & = &
\sum (-1)^n (u^i {\tilde\sigma}^{n-1}+{\tilde\sigma}^n)(\alpha-n) (u^i)^{\alpha-n-1} \,,\\
& = & \sum (-1)^{n+1} {\tilde\sigma}^{n-1} (u^i)^{\alpha-n}\,.
\end{eqnarray*}
So
\[
\left.\left( 
\frac{\partial s^m}{\partial u^i}+\frac{ \partial s^m}{\partial u^{N+1}}
\right)\right|_{u^{N+1}=u^i}=0\,.
\]
The result now follows from the homogeneities of the functions involved.

\endproof

\bigskip

\begin{theorem}

The terms of the coordinates $s^i$ defined above (\ref{eq:newvar}) the
$(g,\Gamma)$ pair depend linearly on the variable $s^{N-1}\,.$

\end{theorem}

\bigskip

\noindent{\bf Proof} In terms of the $u^i$ variables the $(g,\Gamma)$ pair are given by
\begin{eqnarray*}
g^{ij}(u) & = & m^{ij} u^i u^j \,, \\
\Gamma^{ij}_k(u) & = & \delta^j_k m^{ij} u^i
\end{eqnarray*}
where
\[
m^{ij}=\left(
\begin{array}{ccccc}
2-N & 1 & \ldots & 1 & 1 \\
1 & 2-N & \ldots & 1 & 1 \\
\vdots & \vdots & \ddots & \vdots & \vdots \\
1 & 1 & \ldots & 2-N & 1 \\
1 & 1 & \ldots & 1   & N 
\end{array}\right)\,.
\]

\noindent The components of the metric in terms of the $s^i$ coordinates are given
by
\[
{g}^{ij}(s) = \frac{\partial s^i}{\partial u^p}\frac{\partial s^j}{\partial u^q}g^{pq}(u)
\]
and it follows from the symmetry of this equation that the entries will be polynomial in
the new variables. It also follows from this that the degrees of the entries are
\[
{\rm deg}\,{g}^{ij}(s)=\cases{i+j & if $1\leq i\,,j \leq N\,,$\cr
1+j & if $i=N+1$ and $j\neq N+1\,,$ \cr
1+i & if $j=N+1$ and $i\neq N+1\,,$ \cr
2 & if $i=j=N+1\,.$\cr}
\]
The degrees of the terms in the lower right corner of ${g}^{ij}(u)$ are given
schematically below:

\begin{equation}
\left(
\begin{tabular}{c|c|c|c|c}
\multicolumn{4}{c|}{} & \vdots \\
\cline{4-4}
\multicolumn{3}{c|}{} & 2N-2 & N-1 \\
\cline{3-3}
\multicolumn{2}{c|}{} & \multicolumn{1}{c}{2N-2} & 2N-1 & N \\
\cline{2-2}
\multicolumn{1}{c|}{} & \multicolumn{1}{c}{2N-2} & \multicolumn{1}{c}{2N-1} & 2N & N+1 \\
\cline{1-5}
\multicolumn{1}{c}{\ldots} & \multicolumn{1}{c}{N-1} & \multicolumn{1}{c}{N} & N+1 & 2
\end{tabular}
\right)
\end{equation}

\noindent Thus there are only four terms where ${g}^{ij}(s)$ could possibly contain
a term quadratic in $s^{N-1}\,$ (or six terms if $N=2$ or five terms if $N=3\,,$ but
these special cases may be disposed of by direct computation). The result will follow
if it can be shown that these terms contain a factor $s^N\,,$ that is if

\begin{eqnarray*}
{g}^{N-1,N-1}(s) & = & s^N \,\{{\rm polynomial~of~degree~}N-2\,\}\,,\\
{g}^{N,k}(s) & = & s^N \,\{{\rm polynomial~of~degree~}k\,\}\,,\qquad\qquad k=N\,,N-1\,,N-2\,,
\end{eqnarray*}
since the polynomials cannot be quadratic in $s^{N-1}$ without violating the overall degree
of the term.

\bigskip

From these formulae,
\begin{eqnarray*}
{g}^{N,\alpha} & = & \sum_{i,j=1}^N \frac{\partial s^N}{\partial u^i}
\frac{\partial s^\alpha}{\partial u^j}g^{ij} + \sum_{i=1}^N 
\frac{\partial s^N}{\partial u^{N+1}}\frac{\partial s^\alpha}{\partial u^i}g^{N-1,i}\\
& & \sum_{i=1}^N \frac{\partial s^N}{\partial u^i}\frac{\partial s^\alpha}{\partial u^{N+1}}
g^{N-1,i} + \frac{\partial s^N}{\partial u^{N-1}}\frac{\partial s^\alpha}{\partial u^{N-1}}
g^{N-1,N-1}\,.
\end{eqnarray*}
Since
\[
s^N=\prod_{i=1}^N (u^i - u^{N+1})
\]
it follows from Euler's theorem that
\[
\sum_{i=1}^{N+1} u^i \frac{\partial s^N}{\partial u^i} = N s^N
\]
and these may be used to simplify the above. After somewhat tedious calculations
one obtains
\[
\frac{\partial s^N}{\partial u^{N-1}} = \alpha s^N s^\alpha + (1-N) s^N
\left\{
(u^{N+1})^2 \left[\frac{ 	
\frac{\partial s^\alpha}{\partial u^i}+\frac{ \partial s^\alpha}{\partial u^{N+1}} }
{u^i - u^{N+1}}\right] -
u^{N+1} \left[ \frac{\partial s^\alpha}{\partial u^{N+1}} -
\frac{\partial s^\alpha}{\partial u^i} \right]\right\}\,.
\]
The result now follows from the above lemma.

\bigskip

The corresponding result for ${g}^{N-1,N-1}$ is similar and rest on proving, in
a similar manner as above, that

\[
{g}^{N-1,N-1}(s) = (N-1) s^N 
\underbrace{\sum_{i\neq j}
\prod_{ \scriptstyle r=1 \atop \scriptstyle r\neq i\,,j}^{N} (u^r - u^{N+1})\,.
}_{{\rm symmetric~polynomial~in~the~s~variables~of~degree~}(N-2)}
\]

\bigskip

These results show that in the $s^i$-coordinates the metric in linear in the
coordinate $s^{N-1}\,.$ The second half of the proof, showing that $\Gamma^{ij}_k(s)$ is
also linear in $s^{N-1}$ is similar, and follows from the transformation properties
of $\Gamma^{ij}_k\,.$

\endproof

\bigskip

\noindent In what follows this second degenerately flat metric
$\frac{\partial g^{ij}}{\partial s^\bullet}$
will be denoted by $\eta^{ij}\,.$

\bigskip

\begin{example} $N=2\,,M=1\,.$ With these values,

\[
m^{ij}=\left( \begin{array}{ccc}
0 & 1 & 1 \\ 1 & 0 & 1 \\ 1 & 1 & 2\end{array} \right)\,,
\]
and a short computation yields 
\[
g^{ij}(s)=
\left(
\begin{array}{ccc}
2s^2 & s^2(s^1-3s^3) & s^3 (s^1-s^3) \\
s^2(s^1-3s^3) & 2s^2(s^2-s^1s^3+2(s^3)^2) & s^3(2s^2-s^1s^3) \\
s^3(s^1-s^3) & s^3(2s^2-s^1s^3) & 2(s^3)^2
\end{array}
\right)\,.
\]
This is linear in $s^1$ and hence
\begin{eqnarray*}
\eta^{ij}(s) & = & \frac{\partial {g}^{ij}(s)}{\partial s^1} \,, \\
& = & \left(
\begin{array}{ccc}
0 & s^2 & s^3 \\ s^2 & -2 s^2 s^3 & -(s^3)^2 \\ s^3 & -(s^3)^2 & 0
\end{array}
\right)\,.
\end{eqnarray*}

\noindent One may easily introduce degenerate flat coordinates in which the entries
of $\eta^{ij}$ are constant. These flat coordinates are:

\begin{eqnarray*}
t^1 & = & s^1 \,, \\
t^2 & = & \frac{s^2}{s^3} \,, \\
t^3 & = & \log(s^3)
\end{eqnarray*}
and in these coordinates

\[
g^{ij}(t)=\left(\begin{array}{ccc}
2t^2 e^{t^3}   & -2 t^2 e^{t^3} & +t^1 - e^{t^3} \\
-2 t^2 e^{t^3} & +2 t^2 e^{t^3} & -t^1 + e^{t^3} \\
+t^1 - e^{t^3} & -t^1 + e^{t^3} & 2 \end{array} \right)\,.
\]
This is linear in $t^1$ and hence
\[
\eta^{ij}(t) = \left(\begin{array}{ccc}
0 & 0 & +1 \\ 0 & 0 & -1 \\ +1 & -1 & 0
\end{array}\right)\,.
\]

\end{example}

\bigskip

\begin{example} $N=3\,,M=1\,.$ With these values,

\[
m^{ij}=\left( \begin{array}{rrrr} -1 & 1 & 1 & 1 \\ 1 & -1 & 1 & 1 \\ 1 & 1 & -1 & 1 \\
1 & 1 & 1 & +3 \end{array} \right)\,,
\]
and a short computation yields the degenerate metric $g^{ij}(s)\,.$ These components are linear
in $s^2$ and so define a new metric

\begin{eqnarray*}
\eta^{ij}(s) & = & \frac{\partial {g}^{ij}(s)}{\partial s^2} \,, \\
& = & \left(
\begin{array}{cccc}
4 & 0 & 0 & 0 \\
0 & 0 & 2s^3 & 2 s^4 \\
0 & 2 s^3 & -4s^3 s^4 & -2 (s^4)^2 \\
0 & 2s^4 & -2(s^4)^2 & 0
\end{array}
\right)\,.
\end{eqnarray*}

\noindent The degenerate flat coordinates are defined by

\begin{eqnarray*}
t^1 & = & s^1 \,, \\
t^2 & = & s^2 \,, \\
t^3 & = & \frac{s^3}{s^4} \,, \\
t^4 & = & \log(s^4)
\end{eqnarray*}
and in these flat coordinates the original metric metric takes the form $g^{ij}(t)=$

\[
\left(
\begin{array}{cccc}
-(t^1)^2+4t^2 & +6 t^3 e^{t^4} & -6 t^3 e^{t^4} & t^1 - 2 e^{t^4} \\
+6 t^3 e^{t^4} & +4t^1 t^3 e^{t^4}-8t^3 e^{2 t^4}& -4t^1 t^3 e^{t^4}+8t^3 e^{2 t^4} &
+2t^2 - 2 t^1 e^{t^4} + 2 e^{2t^4} \\
-6 t^3 e^{t^4} &-4t^1 t^3 e^{t^4}+8t^3 e^{2 t^4} & +4t^1 t^3 e^{t^4}-8t^3 e^{2 t^4}&
-2t^2 + 2 t^1 e^{t^4} - 2 e^{2t^4} \\
t^1 - 2 e^{t^4}&+2t^2 - 2 t^1 e^{t^4} + 2 e^{2t^4} &-2t^2 + 2 t^1 e^{t^4} - 2 e^{2t^4} & 3
\end{array}
\right).
\]
The entries are linear in $t^2$ and hence one obtains the second metric
\begin{equation}
\eta^{ij}(t)=
\left(
\begin{array}{rrrr}
4 & 0 & 0 & 0 \\
0 & 0 & 0 & 2 \\
0 & 0 & 0 & -2 \\
0 & 2 & -2 & 0
\end{array}
\right)\,.
\label{eq:eta4}
\end{equation}

\end{example}

\bigskip

\section{Degenerate Frobenius manifolds}

The natural geometric setting in which to understand the bi-Hamiltonian structure of
hydrodynamic systems is the Frobenius manifold \cite{D}\,.
One way to defined such manifolds is to construct a function
$F(t^1\,,\ldots\,,t^n)$ such that the associated functions

\[
c_{ijk}=\frac{\partial^3 F}{\partial t^i \partial t^j \partial t^k}
\]
satisfy the following conditions:

\bigskip

\noindent $\bullet$ the matrix $\eta_{ij}=c_{1ij}$ is constant and non-degenerate.
This together with the inverse matrix $\eta^{ij}$ are used to raise and lower indices.
On such a manifold one may interpret $\eta_{ij}$ as a flat metric;

\bigskip

\noindent $\bullet$ the functions $c^i_{jk}=\eta^{ir} c_{rjk}$ defined an associative
commutative algebra with a unity element. This defines a Frobenius algebra on each
tangent space $T^t\M\,.$ This multiplication will be denoted by $u\cdot v\,;$
\bigskip

\noindent $\bullet$ the functions $F$ satisfies a quasi-homogeneity condition, which
may be expressed as

\begin{equation}
\L_E F = d_F\, F + {\{\rm quadratic~terms\}}\,,
\label{eq:homogeneity}
\end{equation}
where $E$ is a vector fields known as the Euler vector field.

\bigskip

\noindent These conditions constitute the Witten-Dijkgraaf-Verlinde-Verlinde (or WDVV)
equations. On such a manifold one may introduce a second flat metric defined by

\begin{equation}
g^{ij}=E(dt^i \cdot dt^j)\,.
\label{eq:intersection}
\end{equation}
This metric, together with the original metric $\eta^{ij}$ define a flat pencil
(i.e. $\eta^{ij}+\lambda g^{ij}$ is flat for all values of $\lambda\,.$ Thus
one automatically obtains a bi-Hamiltonian structure from a Frobenius manifold.
The corresponding Hamiltonians are defined recursively by the formula

\begin{equation}
\frac{\partial^2 h^{(n)}}{\partial t^i\partial t^j}=
c^k_{ij} \frac{\partial h^{(n-1)}}{\partial t^k}\,.
\label{eq:recursion}
\end{equation}
the integrability conditions for this systems are automatically satisfied when
the $c^k_{ij}$ are defined as above.

\bigskip

One basic assumption in this definition is that the metric $\eta_{ij}$ is non-degenerate,
and it follows from this that the bi-Hamiltonian structures are also non-degenerate.
Thus the degenerate bi-Hamiltonian structures obtained in the preceding section cannot
be obtained from this construction. However, one may formulate the new notion of a
degenerate Frobenius manifold in which the corresponding bi-Hamiltonian structures
are degenerate.

\bigskip

Rather than develop the theory of degenerate Frobenius manifolds in full generality
an extended example will be given here based on the study of the hydrodynamic system

\begin{eqnarray*}
u_\tau & = & u(v_x-w_x) \,, \\
v_\tau & = & v(u_x-w_x) \,, \\
w_\tau & = & w(u_x+v_x-2w_x)
\end{eqnarray*}
obtained from the rational Lax function

\[
\L=\frac{(p+u)(p+v)}{p+w}\,.
\]

\noindent The bi-Hamiltonian structure of this system has already been derived
in example 3 in section 4. To recapitulate, in the flat coordinates given by

\begin{eqnarray*}
t^1 & = & u+v-w \,,\\
t^2 & = & \frac{(u-w)(v-w)}{w} \,, \\
t^3 & = & \log w\,,
\end{eqnarray*}
the degenerate metrics which give rise to the degenerate bi-Hamiltonian structures
are:

\begin{eqnarray}
\eta^{ij}(t) & = & \left(\begin{array}{ccc}
0 & 0 & +1 \\ 0 & 0 & -1 \\ +1 & -1 & 0
\end{array}\right)\,, \label{eq:metric1}\\
{g}^{ij}(t) & = & \left(\begin{array}{ccc}
2t^2 e^{t^3}   & -2 t^2 e^{t^3} & +t^1 - e^{t^3} \\
-2 t^2 e^{t^3} & +2 t^2 e^{t^3} & -t^1 + e^{t^3} \\
+t^1 - e^{t^3} & -t^1 + e^{t^3} & 2 \end{array} \right)\,.\label{eq:metric2}
\end{eqnarray}
The first few Hamiltonian densities (suitably normalised)
are given by the formula (\ref{eq:hamdensity}), and in the flat coordinates these become

\begin{eqnarray*}
h^{(1)} & = & t^1 \,, \\
h^{(2)} & = & \frac{1}{2} \left[ (t^1)^2 + 2 t^2 e^{t^3} \right] \,, \\
h^{(3)} & = & \frac{1}{6} \left[ (t^1)^3 + 6 t^1 t^2 e^{t^3} - 3 t^2 e^{2t^3} \right] \,, \\
h^{(4)} & = & \frac{1}{24} \left[
(t^1)^4 + 12 (t^1)^2 t^2 e^{t^3}+ 6 (t^2)^2 e^{2t^3} - 12 t^1 t^2 e^{2t^3} + 4 t^2 e^{3t^3}
\right]\,.
\end{eqnarray*}

\noindent From these and the recursion equation (\ref{eq:recursion}) one may reconstruct
the structure functions $c^i_{jk}$ and verify that they form an commutative and associative
algebra with a unity element. Explicity the structure constants are given by
$c^i_{1j}=\delta^i_j$ and

\[
\left(\begin{array}{ccc}
c^1_{22} & c^2_{22} & c^3_{22} \\
c^1_{23} & c^2_{23} & c^3_{23} \\
c^1_{33} & c^2_{33} & c^3_{33}
\end{array}\right)=\left(\begin{array}{ccc}
0 & 1 & -1/t^2 \\
+e^{t^3} & -e^{t^3} & 0 \\
+t^2 e^{t^3} & -t^2 e^{t^3} & -e^{t^3}
\end{array}\right)\,.
\]
From these structure functions one may raise an index using $\eta^{ij}$ and determine
the Euler vector field from equation (\ref{eq:intersection}). For this example this
vector field is

\[
E=t^1 \frac{\partial~}{\partial t^1}+
t^2 \frac{\partial~}{\partial t^2}+\frac{\partial~}{\partial t^3}\,.
\]

\noindent In addition, the structure functions satisfy the relations

\[
\frac{\partial c^r_{jk}}{\partial t^i} - \frac{ \partial c^r_{ik}}{\partial t^k}=0
\]
and this, together with the symmetry $c^k_{ij}=c^k_{ji}$ enables one to write the
them as

\[
c^i_{jk} = \frac{\partial^2 f^i}{\partial t^j \partial t^k}
\]
for some set of functions $f^i\,.$ For the above structure constants these turn out to
be (up to linear terms),

\begin{eqnarray*}
f^1 & = & \frac{1}{2} (t^1)^2 + t^2 e^{t^3} \,, \\
f^2 & = & \frac{1}{2} (t^2)^2 + t^1 t^2 - t^2 e^{t^3} \,,\\
f^3 & = & t^1 t^3 - t^2 \log t^2 - e^{t^3}\,.
\end{eqnarray*}
At this stage one normally lowers the $i$ index and
use another symmetry to write $c_{ijk}$ as the third derivative of some function $F\,.$
This, however, assumes that the metric $\eta^{ij}$ is invertible, which, for the metric given
by (\ref{eq:metric1}) is not the case. However one may write the $f^i$ as

\[
f^i(t) = \eta^{ij} \frac{\partial F}{\partial t^j} + h^i(t^1+t^2)\,.
\]
Since the matrix $\eta^{ij}$ is of rank 2 it follows it has a non-trivial kernel, so there
exists a non-zero vector $\zeta_i$ such that $\eta^{ij}\zeta_j=0\,,$ and the functions
$h^i$ are functions of the combination $\zeta_i t^i\,,$ which in this example is just
$t^1+t^2\,.$
These functions $h^i$
satisfy
the single constraint $h^1+h^2 = 1/2 (t^1+t^2)^2\,.$
To obtain the above structure functions one possible such $F$ is

\[
F=\frac{1}{2} (t^1)^2 t^3 + t^2 e^{t^3} + \frac{1}{2} (t^2)^2 \log t^2\,,
\]
and
\begin{eqnarray*}
h^1 & = & 0 \,,\\
h^2 & = & \frac{1}{2} (t^1+t^2)^2 \,, \\
h^3 & = & 0 \,.
\end{eqnarray*}
and this satisfies the homogeneity condition (\ref{eq:homogeneity}) with $d_F=2\,.$
There is much freedom in these functions. One may transform, for arbitrary constant $k\,,$
$F$
\[
F\longrightarrow F + k t^3 (t^1+t^2)^2
\]
and the homogeneity property is unchanged. This induced a change in the functions $h^i$
but leaves unchanged the structure functions defining the Frobenius algebra.

\bigskip

From this extended example one may distil the basic properties of a degenerate
Frobenius manifold. One starts with a basic function $F(t^i)$ satisfying some
homogeneity condition and degenerate metric $\eta^{ij}\,,$ the entries of which
are constant in the $t^i$-coordinates. The metric is not related to the third
derivatives of $F\,,$ as for non-degenerate Frobenius manifolds.
The structure functions, which form a Frobenius algebra with a degenerate inner product,
are defined by

\begin{equation}
c^i_{jk} = \eta^{ir} \partial_r \partial_j \partial_k F + \partial_j\partial_k h^i\,,
\label{eq:degstrufunc}
\end{equation}
where the functions $h^i$ are functions which depend on the kernel of the degenerate
matrix $\eta^{ij}\,.$ Thus for degenerate Frobenius manifolds one has a set of extra functions
related to the fact that the matrix $\eta^{ij}$ is not of maximal rank. The
associativity conditions results in a complicated set over-determined partial
differential equations for $F\,,$ the degenerate analogue of the WDVV equations.
One avenue for future research is to develop the concept of
a degenerate Frobenius manifold more axiomatically.

\bigskip

\begin{example} For $N=3\,,M=1$ the metrics $g^{ij}(t)$ and $\eta^{ij}(t)$ have been calculated
in example 4. One may repeat the calculations above and obtain

\bigskip

\noindent$\bullet$ Euler vector field:

\[
E=\frac{t^1}{2} \frac{\partial~}{\partial t^1} + t^2 \frac{\partial~}{\partial t^2} +
t^3 \frac{\partial~}{\partial t^3} + \frac{1}{2} \frac{\partial~}{\partial t^4}\,;
\]

\noindent$\bullet$ Prepotential $F\,:$

\[
F= \frac{1}{8} (t^1)^2 t^2 + \frac{1}{4} (t^2)^3 - \frac{1}{192} (t^1)^4 + \frac{1}{2}
t^1 t^3 e^{t^4} - \frac{1}{4} t^3 e^{2 t^4} - \frac{1}{4} (t^3)^2 \log t^3\,;
\]

\noindent$\bullet$ associated non-zero potentials $h^i\,:$

\[
h^3 = \frac{1}{2} (t^2+t^3)^2\,.
\]

\noindent From these, and the constant matrix $\eta^{ij}$ given by (\ref{eq:eta4}), one
may construct a degenerate Frobenius algebra with structure functions given by equation
(\ref{eq:degstrufunc}) and second degenerate flat metric given by (\ref{eq:intersection}).

\end{example}

\bigskip

In section 4 the bi-Hamiltonian structures were shown to exist for arbitrary $N$ but
$M=1\,.$ It is clear that the ideas will generalize to arbitrary $M\,,$ and hence
to degenerate Frobenius manifolds for arbitrary $N$ and $M\,.$ The following example
is for $N=3\,,M=2\,.$
\bigskip

\begin{example} For $N=3\,,M=2$ the flat coordinates are defined by the expansion

\[
\L=p + t^1 + \frac{ t^2 e^{t^4} }{p+e^{t^4} } + \frac{ t^3 e^{t^5} }{p+e^{t^5} }\,.
\]

\noindent In these coordinates

\[
\eta^{ij}(t) = \left(
\begin{array}{rrrrr}
0 & 0 & 0 & 1 & 1 \\
0 & 0 & 0 & -1 & 0 \\
0 & 0 & 0 & 0 & -1 \\
1 & -1 & 0 & 0 & 0 \\
1 & 0 & -1 & 0 & 0
\end{array}\right)
\]

\noindent and the Frobenius data is:

\bigskip

\noindent$\bullet$ Euler vector field:

\[
E=t^1 \frac{\partial~}{\partial t^1} + t^2 \frac{\partial~}{\partial t^2} +
t^3 \frac{\partial~}{\partial t^3} + \frac{\partial~}{\partial t^4} +
\frac{\partial~}{\partial t^5} \,;
\]

\noindent$\bullet$ Prepotential $F\,:$

\begin{eqnarray*}
F & = & +t^2 e^{t^4} + t^3 e^{t^5} + \frac{1}{2} t^2 t^3 \log( e^{t^4}-e^{t^5} )^2+
\frac{1}{2} \left( (t^2)^2 \log t^2 + (t^3)^2 \log t^3 \right) \\
& = & -\frac{1}{2} t^2 t^4 (2 t^1 + t^2 + 2 t^3) - \frac{1}{2} t^3 t^5 (2 t^1 + 2 t^2 + t^3)\,;
\end{eqnarray*}

\noindent$\bullet$ associated non-zero potentials $h^i\,:$

\[
h^1 = \frac{1}{2} (t^1 +t^2+t^3)^2\,.
\]

\noindent From this data the Frobenius algebra structure functions given by equation
(\ref{eq:degstrufunc}) and the second degenerate flat metric given by (\ref{eq:intersection}).

\end{example}

\bigskip

\noindent The form of these results suggest the following:

\begin{conjecture} The metric given in Theorem 7 is linear in the coordinate $s^{N-M}\,,$
where the coordinates $s^i$ are defined in terms of the expansion of the rational
Lax function

\[
\L = p^{N-M} + s^1 p^{N-M-1} + \ldots + s^{N-M}+\frac{s^{N-M+1}}{p+ s^{N+1}}+
\ldots + \frac{s^{N}}{p+ s^{N+M}}\,.
\]

\noindent Moreover, there exist flat coordinates $t^i$ such that the variables
$s^i$ are polynomial functions of the variables
$t^1\,,\ldots\,,t^N\,,e^{t^{N+1}}\,,\ldots e^{t^M}\,,$
and in which the entries $\eta^{ij}(t)$ are all constants.

\end{conjecture}

\noindent One would hope to be able to modify the results of \cite{DZ} to prove this
conjecture; the vanishing of the determinants of the metrics means that the results
cannot be used directly. One should be able to modify the Gauss-Manin equations for
the flat coordinates to include these degenerate examples.

\bigskip

\section{Comments}

One notable difference between the bi-Hamiltonian structure of the hierarchies consider here,
these being multi-component generalizations of Toda and Benney hierarchies \cite{FS},
and the bi-Hamiltonian structures of dispersionless KP-type hierarchies is the degeneracy
of the structures. The dispersionless KP-type hydrodynamic systems involve rational
such as (see, for example, those in \cite{AK})

\[
\L=\frac{1}{2} p^2 + S(x,t) + \frac{P(x,t)}{p-Q(x,t)}
\]
and Lax equation similar to equation (\ref{eq:lax}), but with Poisson bracket

\[
\{ f,g\}_{PB} = (\partial_p f \partial_x g - \partial_x f \partial_p g)\,.
\]
The bi-Hamiltonian structure of these equations are not degenerate \cite{D,AK}. These
rational Lax functions may be considered as a reduction of an infinite component
Lax function $\L=\sum_{i=-\infty}^N s^i p^i$ and it may be of interest to see how
constraining the resulting Hamiltonian structures results in the degenerate structures
studied here.

\bigskip

The existence of a non-trivial Casimir for these systems is of interest. One possible reduction
of these systems is to restrict the dynamics to the surface given by

\[
{\cal C} = {\rm constant}\,,
\]
for example the $(N=2\,,M=1)$ system

\begin{eqnarray*}
u_\tau & = & u (v_x-w_x)\,, \\
v_\tau & = & v (u_x-w_x)\,, \\
w_\tau & = & w (u_x+v_x-2w_x)
\end{eqnarray*}
when restricted to the surface $w=uv$ results in the system

\begin{eqnarray*}
u_\tau & = & u [ (1-u) v_x - v u_x ] \,, \\
v_\tau & = & v [ (1-v) u_x - u v_x ] \,.
\end{eqnarray*}
How the Hamiltonian structure behaves under such a constraint is unknown.
For non-degenerate Hamiltonian structures one may use the result of
Ferapontov \cite{Fera}, though this work would need to be generalised to include
degenerate Hamiltonian structure such as those considered here. More generally,
one may restrict the above system to the surface $w=u v f(x)$ for some arbitrary
function $f(x)$ (i.e. ${\cal C}=f^{-1}\,.$). This results in the
system

\begin{eqnarray*}
u_\tau & = & u [ (1-u) v_x - v u_x ] - u^2 v f^\prime(x)\,, \\
v_\tau & = & v [ (1-v) u_x - u v_x ] - v^2 u f^\prime(x)\,,
\end{eqnarray*}
an example of inhomogeneous hydrodynamic system with specific $x$-dependence. It may
also be possible to obtain the Hamiltonian structure of these systems \cite{Polyak}\,.

\bigskip

The idea of a degenerate Frobenius manifold requires further elucidation. One complicating
factor is that for a degenerate structure the transformation which reduces the components
of the metric to constants will not, in general, reduce all of the components
$\Gamma^{ij}_k$ to zero \cite{G}. The systems in this paper are special in this respect since in
flat coordinates the components of $\Gamma^{ij}_k$ are automatically zero, which is
not the generic suituation; the systems here are doubly degenerate.

\bigskip

Finally, this paper has only dealt with dispersionless systems. For polynomial
Lax equations one has discrete counterparts, the simplest example being the
Toda Lattice

\begin{eqnarray*}
S_{n,\tau} & = & P_n - P_{n+1} \,, \\
P_{n,\tau} & = & P_n ( S_{n-1} - S_n)
\end{eqnarray*}

\noindent which reduces to (\ref{eq:example}) in the continuum limit; the lattice
variable becoming the continuous variable $x\,.$ The bi-Hamiltonian structure of
such systems have been studied in \cite{K}. Indeed, the structures obtained here
could also be derived by taking certain limits of those structures, if they where
known explicitly for arbitrary $M$ and $N\,.$ How to extend these results to rational
discrete systems is unclear. One approach would be to use the ideas in \cite{EOR}, which
deals with the interpretation of the inverse operator $(e^\partial+u)^{-1}\,,$ or the
ideas of \cite{EYY} where one would consider term-by-term deformation of the
underlying dispersionless system.

\bigskip

\section*{Acknowledgments}
I would like to thank Prof. A.P. Fordy for informing me that the result of
Dubrovin \cite{D} was a special case of a result of Magri \cite{Magri}\,.

\end{document}